\documentclass{iopart}
\usepackage{graphicx} 
\usepackage{rotating}
\usepackage{cite}

\begin{document}

\title{Substituting the main group element in cobalt - iron based Heusler alloys:
Co$_2$FeAl$_{1-x}$Si$_x$.}

\author{Gerhard H. Fecher and Claudia Felser}
\address{Institut f\"ur Anorganische Chemie und Analytische Chemie,\\
         Johannes Gutenberg-Universit\"at Mainz,
         D-55099 Mainz, Germany}
\ead{fecher@uni-mainz.de}

\date{\today}

\begin{abstract}

This work reports about electronic structure calculations for the Heusler compound Co$_2$FeAl$_{1-x}$Si$_x$. Particular emphasis was put on the role of the main group element in this compound. The substitution of Al by Si leads to an increase of the number of valence electrons with increasing Si content and may be seen as electron-doping. Self-consistent electronic structure calculations were performed to investigate the consequences of the electron doping for the magnetic properties. The series Co$_2$FeAl$_{1-x}$Si$_x$ is found to exhibit half-metallic ferromagnetism and the magnetic moment follows the Slater-Pauling rule. It is shown that the electron-doping stabilises the gap in the minority states for $x=0.5$.

\end{abstract}



\submitto{\JPD}

\section{Introduction}

Heusler compounds \cite{Heu03,HSH03,Heu04} have attracted scientific and technological interest for their potential use as materials for magneto-electronic devices. Reason is the exceptional electronic structure found in many of those compounds, in particular in those based on cobalt. They exhibit a complete spin polarisation at the Fermi energy ($\epsilon_F$), that means they behave like a metal for electrons of one spin direction and like an insulator for the other. K{\"u}bler \etal \cite{KWS83} recognised that the minority-spin density at the Fermi energy nearly vanishes in the Heusler compounds Co$_2$MnAl and Co$_2$MnSn. The authors concluded that this should lead to peculiar transport properties in these compounds because only the majority density contributes. Materials with a complete spin polarisation at $\epsilon_F$ are called half-metallic ferromagnets \cite{GME83}, even though there do exist more complicated cases as classified in Ref.~\cite{CVB02}.

The Heusler compounds are usually ternary 2-1-1 compounds. They consist for the most part of two transition metals (X$_2$, Y) and one main group (Z) element crystallising in the $L2_1$ structure (space group $F\:m\overline{3}m$). Besides ternary X$_2$YZ compounds, there exist also large assortments of substitutional quaternary alloys of the type X$_2$Y$_{1-x}$Y'$_x$Z or X$_2$YZ$_{1-x}$Z'$_x$.

One of the early substitutional series that attracted interest as potential material for magneto-electronics was Co$_2$Cr$_{1-x}$Fe$_x$Al \cite{EFV03,BFJ03,FKW05,WFK06b}. Drawback of this series is that it is hard to be stabilised in the $L2_1$ structure. Mostly a mixture of atoms in Y and Z positions is observed leading to $B2$-like disorder \cite{KUK04}. However, the disorder destroys the half-metallic properties \cite{WFK06b}. Recently, the series of Heusler alloys Co$_2$Mn$_{1-x}$Fe$_x$Si has attracted particular interest because it exhibits the $L2_1$ order over the whole range of $x$ \cite{BFK06}. The Curie temperatures of the end members are 985~K \cite{FSI90,BNW00} and 1100~K \cite{WFK05,WFK06a} for the Mn and Fe containing compounds, respectively. The end members of the series Co$_2$Mn$_{1-x}$Fe$_x$Si, that are the purely Mn or Fe containing compounds, have been used for fabrication of magnetic tunnel junctions \cite{IOM06,OSN06}. The tunnel magneto-resistance (TMR) ratios of 159\% in the Mn compound at low temperature and 41\% in the Fe compound at room temperature suggest that still an improvement in the materials is necessary for successful use in devices, in particular with respect to their temperature behaviour. Recently, Tekuda \etal \cite{TIM06} reported about tunnel junctions build from the iso-electronic compound Co$_2$FeAl$_{0.5}$Si$_{0.5}$. The junctions exhibited TMR ratios of 76\% at 300~K and 106\% at 5~K for the $B2$ structure while that with $L2_1$ structure showed 51\% and 78\% at 300~K and 5~K, respectively \footnote{The TMR ratio is 175\% at 300~K for optimised junctions with $L2_1$ structure, private communication by K.~Inomata (Tsukuba, Japan).}. These values of the TMR ratio are larger than the ones found using pure Co$_2$FeAl or Co$_2$FeSi electrodes.

The temperature stability of the minority gap is one of the main challenging questions for materials to be used in applications. From the viewpoint of the electronic structure, several different effects may destroy the half-metallicity at finite temperatures, depending on the situation of $\epsilon_F$. At $T>0$, quasi-particle states - occurring close to the minority band edges - may be induced in the gap \cite{CAK06}, for example by magnon excitation. In particular, a spin-rotation \cite{SDo02} may destroy or at least reduce the size of the gap. This has the effect that $\epsilon_F$ - being initially situated at the top of the valence band - does not fall any longer inside of the gap at elevated temperature. If $\epsilon_F$ is located at the bottom of the conduction band then the half-metallicity will be immediately lost for $T>0$ due to the occupation of minority states through the Fermi-Dirac distribution. (Note: this effect cannot appear the same way for $\epsilon_F$ being situated at the top of the valence band, as there are no states to be occupied thermally inside of the gap but only above.) Finally, the lattice parameter and defect densities will be changed at elevated temperatures resulting also in changes of the electronic structure. An already small increase of the lattice parameter may be able to destroy the half-metallicity if the Fermi energy was initially at $T=0$ located close to one of the edges of the minority gap. At the same situation, a smearing of the states close to $\epsilon_F$ by an increase of the defects with temperature above 0~K, will also destroy the half-metallic character. Taking all those facts together, it is expected that a location of $\epsilon_F$ close to the middle of the minority gap will result in the most robust half-metallicity, supposed the gap is not too small ($\approx 1$~eV).

Low magnetic moment compounds like Co$_2$CrAl exhibit a variety of majority $d$-bands crossing the Fermi energy in rather all directions of the Brillouin zone. The high magnetic moment Co$_2$YZ compounds ($m>4\mu_B)$ exhibit only few, strongly dispersing majority $d$-bands crossing $\epsilon_F$ mainly along high symmetry directions. These few bands may be in favour for coherent tunnelling \cite{NTI06}. Finally, a mixture of the $3d$ elements on the Y position will cause different localised moments on different sites. This might lead to instabilities of the half-metallic character. All those aspects, fixing the $d$-state element on the Y position and thus the kind of localised moment, location of the Fermi energy with respect to the gap in the minority states, and simplicities of the majority bands crossing it, were of prime importance for selection of the Co$_2$FeAl$_{1-x}$Si$_x$ series for the present study.

\section{Computational details}
\label{sec:CD}

The electronic structure of the series of alloys was calculated by means of the full potential linearised augmented plane wave (FLAPW) method. For Co$_2$FeAl$_{1-x}$Si$_x$, the calculations were carried out using the FLAPW method as implemented in {\scshape Wien}2k provided by Blaha \etal \cite{BSS90,BSM01}. The exchange-correlation functional was taken within the generalised gradient approximation (GGA) in the parameterisation of Perdew \etal \cite{PBE96}. In addition, the LDA$+U$ method \cite{AAL97} was used to respect on-site correlation at the $3d$ transition metals. It should be mentioned that the $+U$ was used here together with the GGA rather than the LSDA parameterisation of the exchange-correlation functional. However, no significant differences were observed using either of these parameterisations. In {\scshape Wien}2k, the effective Coulomb-exchange parameter $U_{eff}=U-J$ is used, where $U$ is the Coulomb part and $J$ is the exchange part. The use of $U_{eff}$ suppresses multipole effects. That means, it neglects the non-spherical terms in the expansion of the Coulomb interaction. In particular, the values for $U_{eff}$ were set to $U_{Co}=0.14$~Ry, and $U_{Fe}=0.132$~Ry, independent of the Si concentration. These values are able to explain the the magnetic moment in Co$_2$Mn$_{1-x}$Fe$_x$Si over the whole range of Fe concentration, as was found in previous calculations \cite{BFK06}. $U_{Co}$ and $U_{Fe}$ are close to the values for the Coulomb interaction $U_{dd}$ for $d$ electrons in the elemental $3d$ transition metals reported in Ref.~\cite{BSa89}. Finally, a $25\times25\times25$ point mesh was used as base for the integration in the cubic systems resulting in 455 $k$-points in the irreducible wedge of the Brillouin zone. No noticeable changes in the precision of the magnetic moments or in the position of the Fermi energy were observed if comparing to a smaller $20\times20\times20$ mesh. The energy convergence criterion was set to $10^{-5}$~Ry and simultaneously the criterion for charge convergence to $10^{-3}$ electrons. This combination resulted in final values being about one order of magnitude lower for both criteria.

The properties of the pure Al or Si containing compounds were calculated in $F\:m\overline{3}m$ symmetry using the lattice parameter ($a_{Al}=5.706$~\AA, $a_{Si}=5.633${~\AA}) as found from a structural optimisation. The values are close to the experimental ones: $a_{Al}=5.727$~\AA and $a_{Si}=5.64$~\AA. The larger deviation for the Al compound may be caused by the fact that this compound does frequently not have an ordered $L2_1$ structure in experiments. Following Vegard's law, a linear variation of $a$ was assumed for the mixed compounds. All muffin tin radii were set to nearly touching spheres. The mixed compounds with $x=1/4$ and $3/4$ were calculated in $P\:m\overline{3}m$ and for $x=1/2$ in $P\:4/mmm$ symmetry, similar as for the mixed Cr-Fe compounds as reported in \cite{FKW05}.

\section{Results and Discussion}

\begin{figure}
\includegraphics[width=6cm]{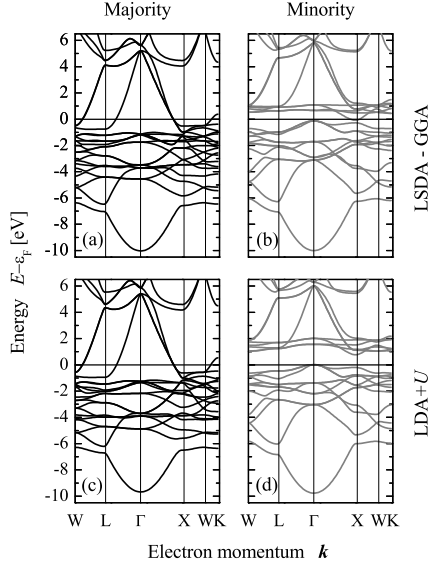}
\caption{Spin resolved band structure of Co$_2$FeAl. \newline
         Compared are the band structures calculated in the LSDA-GGA (a,b) and
         the LDA$+U$ (c,d) approaches.}
\label{fig_1}
\end{figure}

In the following the electronic and magnetic structure of the series Co$_2$FeAl$_{1-x}$Si$_x$ will be discussed. The end members Co$_2$FeAl and Co$_2$FeSi were already presented in detail in previous work. In Refs.~\cite{WFK05,KFF06} it was shown, that it is necessary to include the on-site correlation in the calculations for Co$_2$FeSi in order to explain the experimental data and to find the half-metallic ground state. Details of the band structure for that system are found in Refs.~\cite{BFK06,KFF06}. Already in pure LSDA-GGA calculations, Co$_2$FeAl became a half-metallic ferromagnet \cite{FKW05}. It can be expected, however, that on-site correlation plays also an important role in the Al containing compound if it does in the case of Si. Therefore, the electronic structure of Co$_2$FeAl was recalculated using the LDA$+U$ method as described in section \ref{sec:CD}. Figure~\ref{fig_1} compares the spin resolved band structure of Co$_2$FeAl calculated in the LSDA-GGA and the LDA$+U$ approach.

From Figure~\ref{fig_1}, it is seen that the inclusion of $U_{eff}$ in the calculation does not cause pronounced changes of the majority bands. Even the flat band at about -4~eV below the Fermi energy is shifted by only 200~meV to higher binding energies. This is remarkable as this band is mainly responsible for the localised moment at the Fe atom. The major impact of the Coulomb parameter is on the minority bands and in particular on their unoccupied part. The gap is clearly opened up and the flat, lowest conduction bands at the $\Gamma$-point are shifted by about 1~eV to higher energies.

\begin{figure}[ht]
\includegraphics[width=6cm]{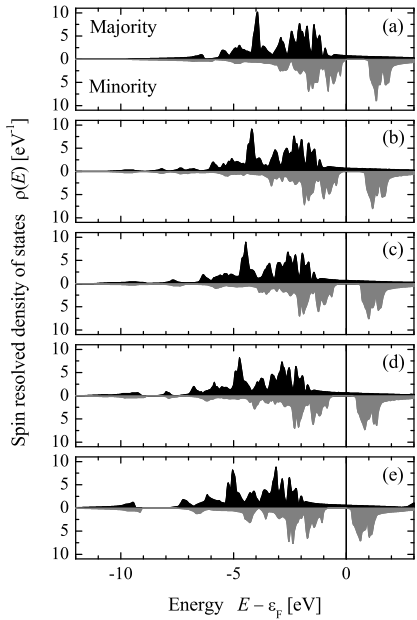}
\caption{Spin resolved density of states of Co$_2$FeAl$_{1-x}$Si$_x$. \newline
         The panels (a, ... , e) show - from top to button - the DOS with increasing amount of Si
         for $x=0, 0.25, 0.5, 0.75$, and 1. The DOS is calculated using LDA$+U$.}
\label{fig_2}
\end{figure}

Figure \ref{fig_2} compares the spin resolved density of states for the complete series Co$_2$FeAl$_{1-x}$Si$_x$ The calculations were carried out using LDA$+U$. For $x=0$ the low lying $s$-states are found at energies below -6~eV. With increasing Si content a new group of $s$-states appears that is found for $x=1$ at energies below -9~eV. In the range of intermediate Si concentration both groups appear. Those low lying $s$-states are separated from the $p$ and $d$-states by the Heusler-typical hybridisation gap. This gap is considerably larger in the Si compound ($\approx 1.5$~eV) compared to the Al compound ($\approx 0.5$~eV) indicating the stronger hybridisation. This stronger hybridisation makes the Si rich alloys more stable compared to the Al rich part. The $p$-states are in all cases found at the bottom of the high lying valence bands above that gap. With increasing $x$, both - the majority as well as the minority channel - do not exhibit pronounced changes of the $d$-state derived densities. Their general shape stays rather unaffected. However, the $d$-band width increases from about 5.2~eV to 6.7~eV with increasing Si content. At the same time, the high majority density of the localised $d$-states found at -4~eV in the Al compound shifts to -5~eV in the Si compound. Fixing the Fermi energy in the minority gap at the position where it is found in the Al compound would thus lead to a rather unphysical enlargement of the exchange splitting. However, the shift of the majority density is compensated by a shift of the minority density with increasing Si content to lower energies. As a result, one observes a virtual movement of $\epsilon_F$ through the gap in the minority states. Throughout the whole series from $x=0$ to 1, the band gap about $\epsilon_F$ is clearly revealed in the minority density. The properties of the gap are further discussed after a brief discussion of the magnetic moments.

\begin{figure}[ht]
\includegraphics[width=8cm]{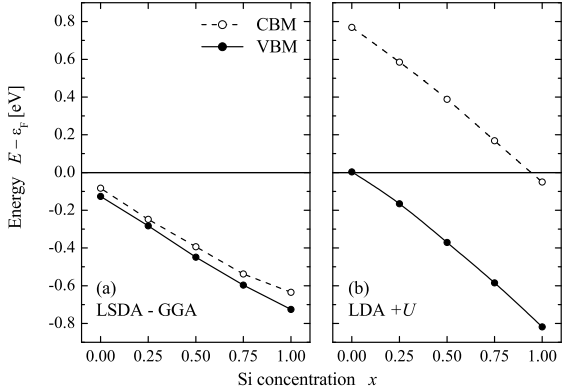}
\caption{The gap in the minority states of Co$_2$FeAl$_{1-x}$Si$_x$. \newline
         Compared are the positions of the valence band maximum (VBM) and the
         conduction band minimum (CBM) as calculated by means of LSDA-GGA in (a) and
         LDA$+U$ in (b).}
\label{fig_3}
\end{figure}

The magnetic properties are compared in Tabel~\ref{tab_1}. The nearly half-metallic state found for Co$_2$FeAl already in the LSDA-GGA calculation results in a nearly integer spin magnetic moment (deviation smaller than $10^{-3}$). This reflects the well-expected fact that a magnetic moment being compatible to the Slater-Pauling rule ($m=(N_v-24)\mu_B$, with $N_v$ being the number of valence electrons in the primitive cell containing four atoms \cite{Kue00,FKW06}) must not result unambiguously in a half-metallic state. The substitution of Al by Si results in a pronounced deviation from the Slater-Pauling-like behaviour, in the LSDA-GGA approach. This is clear from the shift of the minority band-gap away from the Fermi energy in the LSDA-GGA calculations (see Fig.~\ref{fig_3}(a)).

\begin{table}[ht]
\centering
\caption{Total magnetic moments of ordered Co$_2$FeAl$_{1-x}$Si$_x$. \\
         All moments were calculated for the given super-cells.
         Their values are in $\mu_B$ and respect 4 atoms in the cell for easier comparison.}
     \begin{tabular}{l|c|cc}
        compound             & $x$   & GGA  & LDA$+U$ \\
        \noalign{\smallskip}\hline\noalign{\smallskip}
        Co$_2$FeAl           & 0     & 5.0   & 5.0  \\
        Co$_8$Fe$_4$Al$_3$Si & $1/4$ & 5.2   & 5.25 \\
        Co$_4$Fe$_2$AlSi     & $1/2$ & 5.37  & 5.5  \\
        Co$_8$Fe$_4$AlSi$_3$ & $3/4$ & 5.49  & 5.75 \\
        Co$_2$FeSi           & 1     & 5.53  & 6.0  \\
        \noalign{\smallskip}\hline
    \end{tabular}
    \label{tab_1}
\end{table}

Other than in pure LSDA-GGA, the LDA$+U$ calculations reveal clearly a linear dependence of the spin magnetic moment $m(x)$ on the Si concentration $x$ and thus on the number of valence electrons. The small deviation from a half-metallic state for the $x=0$ and 1 compounds seen in Fig.~\ref{fig_3}(b) does not lead to a discernible deviation from the expected integer moments of 5~$\mu_B$ or 6~$\mu_B$, respectively. This only small deviation is caused by the fact that only very few states contribute to the minority density close to $\epsilon_F$ (see Fig.~\ref{fig_2}). Finally, it has to be noted that the LSDA-GGA calculations do not reflect the experimentally found magnetic moment of 6~$\mu_B$ for Co$_2$FeSi, either at the optimised or at the experimental lattice parameter. It was carefully checked that this effect is not caused by a missing orbital magnetic moment. Using LSDA together with spin-orbit coupling and Brook's orbital polarisation resulted in $m_{s}=5.536\mu_B$ and $m_{l}=0.189\mu_B$. This means that $m_{tot}=5.725\mu_B$ is still far below the value determined by magnetometry (this magnitude of the total magnetic moment $m_{tot}$ was observed using {\scshape Wien}2k as well as the relativistic {\scshape Munich}-SPRKKR, not reported here.). Only the LDA$+U$ scheme was able to reflect the value in the correct order, independent whether spin-orbit interaction was respected or not.

Figure \ref{fig_3} compares the behaviour of the gap in the minority states of Co$_2$FeAl$_{1-x}$Si$_x$. Shown are the extremal energies of the states involving the minority band gap that are the accompanied valence band maximum and conduction band minimum. In the LSDA-GGA approach, the small gap in the minority states is moved away from the Fermi-energy with increasing Si content and the half-metallicity becomes destroyed. Using LDA$+U$, the gap has a nearly constant width of 760~meV over the complete series from $x=0$ to 1. From Figure \ref{fig_3}, it is seen that the end-members are just at the borderline to half-metallic ferromagnetism. Starting from $x=0$, the Fermi energy moves from the top of the valence band to the bottom of the conduction band at $x=1$. For Co$_2$FeAl$_{0.5}$Si$_{0.5}$, the Fermi energy is located close to the middle of the band gap in the minority states. It should be noted that the values for $U_{eff}$ used here are the borderline cases for the half-metallic ferromagnetism over the complete series Co$_2$FeAl$_{1-x}$Si$_x$. Small variations of the Coulomb parameter $U_{eff}$ will change the behaviour of the density of states as reported in Refs.~\cite{WFK05,KFF06} for Co$_2$FeSi. About 10\% higher or lower values will make one of the end members a clear half-metal and destroy at the same time the half-metallic character of the other one. From the theoretical point of view, the compounds Co$_2$FeAl and Co$_2$FeSi are thus very unstable half-metallic ferromagnets, if at all.

\section{Summary and Conclusions}

The electronic structure of the substitutional series of the quaternary Heusler compounds Co$_2$FeAl$_{1-x}$Si$_x$ was investigated by means of band structure calculations using the LDA and LDA$+U$ approximations. It was found that the Co$_2$FeAl$_{1-x}$Si$_x$ series of compounds exhibits half-metallic ferromagnetism if using the LDA$+U$ scheme. Moderate Coulomb-interaction parameters of less than 2~eV were used. For the two end-members, Co$_2$FeAl and Co$_2$FeSi, the Fermi energy is close to the band edges of the minority states. The high densities at those band edges make the half-metallic character of both compounds rather unstable at finite temperatures above 0~K. This might be one reason explaining the low tunnelling magneto resistance ratio found in those compounds at room temperature. For $x\approx0.5$, the calculations predict that the Fermi energy is located in the middle of the gap of the minority states. This behaviour will make Co$_2$FeAl$_{0.5}$Si$_{0.5}$ stable against temperature variations as discussed in the introduction. Experiments were started to verify over what range of compositions the series Co$_2$FeAl$_{1-x}$Si$_x$ crystallises in the required $L2_1$ structure and to find the most stable half-metallic ferromagnet in this series.

In summary, it was shown that the variation of the main group element in Heusler compounds is a strong tool in order to tune their physical properties.

\bigskip
\noindent{\bf Acknowledgment :}\newline

We thank K.~Inomata  (Tsukuba, Japan) for providing his data before publication. The authors are very grateful to P.~Blaha ({\scshape Wien}2k) and H.~Ebert ({\scshape Munich}-SPRKKR) and their groups for development and providing the computer codes. This work is financially supported by the Deutsche Forschungs Gemeinschaft (project TP7 in research group FG 559).

\newpage
\bibliographystyle{unsrt}

\end{document}